\documentstyle[12pt]{article}
\newcounter{num}
\begin{document}
\baselineskip 14pt plus 2pt
\vspace{6cm}

\hspace{9cm}TECHNION-PH-96-17
\vspace{0.5cm}

\begin{center}
THE
$s \rightarrow d \gamma$
TRANSITION IN KAON AND HYPERON DECAYS$^{\ast}$ \\[0.50cm]

Paul Singer \\
Department of Physics, Technion-Israel Institute of Technology \\
Haifa, Israel
\end{center}
\vspace{6cm}

We survey the possibilities of detecting the
$s \rightarrow d \gamma$
transition in kaon and hyperon radiative decays. In the more frequent
decays, like
$K^+ \rightarrow \pi^+ \pi^o \gamma$,
$K^+ \rightarrow \pi^+ \ell^+ \ell^-$,
the short-distance contribution is obscured by various long-distance
transitions. Among the hyperon radiative decays,
$\Xi^- \rightarrow \Sigma^- \gamma$
and
$\Omega^-  \rightarrow \Xi^- \gamma$
are the leading candidates for providing a window to the short-distance
$s \rightarrow d \gamma$.
The long-distance
$s \rightarrow d \gamma$
transitions are also analyzed; their magnitude depends on the size of
the deviation from a sum-rule among couplings of vector mesons to
photons. The measurement of
$\Omega^-  \rightarrow \Xi^- \gamma$
could provide information on the relative size of the short and
long-distance contributions to the magnetic component of
$s \rightarrow d \gamma$.
\vspace{1cm}

---------------------------------------- \\

$^{\ast}$ Lecture presented at the Workshop on K-Physics, Orsay, France,
30 May - 4 June, 1996.
\pagebreak

\baselineskip 22pt plus 2pt
\noindent {\bf 1. \ Introduction} \\

Flavour changing single quark
$Q \rightarrow q+ \gamma$
transitions, induced by loop diagrams, are a basic feature of the
electroweak standard model$^1$. The calculation$^2$ of the amplitude
of such transitions from the electroweak model needs to be complemented
by the inclusion of QCD corrections$^3$. It is of obvious interest to
identify physical processes in which these transitions contribute
significantly. Such processes would test the standard model at the
one loop level, as well as the procedure of administering to it the QCD
corrections and then possibly provide windows to physics beyond the
standard model.

Although the
$s \rightarrow d\gamma$
transition was the first to be investigated$^{1,2,4,5}$ in detail, with
the aim of relating it to physical radiative processes of kaons and
hyperons, it is the
$b \rightarrow s\gamma$
transition$^6$
which has conquered the limelight during recent years. Since it was
shown$^7$
that the enhancement provided by the QCD corrections to the
$b \rightarrow s\gamma$
transition brings the inclusive
$B \rightarrow X_s \gamma$
and the exclusive
$B \rightarrow K^{\ast} \gamma$
modes into the realm of possible detection, the radiative penguin decays
of the b-quark have received considerable attention. This has resulted
in the calculation of the gluonic corrections fully to the leading
order$^8$ and partially at the next-to-leading order$^9$. The recent
measurements by the CLEO collaboration of the inclusive rate$^{10}$
$Br (B \rightarrow X_s \gamma) = (2.32 \pm 0.57 \pm 0.35) \times
10^{-4}$
and of the exclusive decay$^{11}$
$Br (B \rightarrow K^{\ast} \gamma) = (4.5 \pm 1.5 \pm 0.9) \times
10^{-5}$
are in agreement with the theoretical expectations. An
analysis$^{12}$ of the lowest order prediction for the
inclusive mode gives
$Br (B \rightarrow X_s \gamma)^{th} = (2.8 \pm 0.8) \times
10^{-4}$,
with the error due mainly to the uncertainty in the choice$^{13}$
of the renormalization scale. The completion of the next-to-leading
order calculation will reduce significantly the theoretical uncertainty.

It is our purpose to review here the present status of the
$s \rightarrow d \gamma$
transition and to analyze the possibility of its detection.
Before turning to the main topic, we remark that the contribution
of the
$c \rightarrow u \gamma$
transition in charm decays has  been analyzed$^{14}$ recently in detail.
It turns out that the electroweak
$c \rightarrow u \gamma$
QCD uncorrected transition has a minuscule branching ratio of about
$10^{-17}$
(vs.
$10^{-4}$
in the
$b \rightarrow s \gamma$
case). The inclusion of gluonic corrections raises dramatically this
figure to nearly
$10^{-11}$;
however, this is still extremely small and as a result the domain of
weak radiative decays of charmed particles is expected to be dominated
by various long distance contributions.
\vspace{0.5cm}

\noindent {\bf 2. \ The Flavour-Changing
$Q \rightarrow q \gamma$
Transition in the Standard Model}  \\

The
penguin amplitude for the transition of heavy quark
$Q$
to a light quark $q$, with quarks
$Q,q$
on the mass-shell is given$^{1,2}$ by
\begin{eqnarray}
&& \Gamma^{(Q \rightarrow q \gamma)}_{\mu}
= \frac{eG_F}{4 \pi^2 \sqrt{2}}
\sum_{\lambda}
V^{\ast}_{\lambda Q}
V_{\lambda q}
\bar{u} (q)
[F_{1,\lambda}(k^2)(k_{\mu} k\hspace{-0.22cm}/ \nonumber \\
&& - k^2 \gamma_{\mu})
\frac{1-\gamma_5}{2}
+ F_{2,\lambda}(k^2) i\sigma_{\mu \nu} k^{\nu}
(m_Q
\frac{1+\gamma_5}{2}
+ m_q
\frac{1-\gamma_5}{2})] u(Q) \ .   
\end{eqnarray}

$F_1 = \sum_{\lambda}
V^{\ast}_{\lambda Q}
V_{\lambda q}
F_{1,\lambda}$
and
$F_2 = \sum_{\lambda}
V^{\ast}_{\lambda Q}
V_{\lambda q}
F_{2,\lambda}$
are the charge radius and magnetic form factors respectively and
$V_{ab}$
are CKM matrices. They were firstly calculated in the electroweak model
by Inami and Lin$^1$,
$F_{1,\lambda}$
and
$F_{2,\lambda}$
giving the contribution of each internal quark in the loop. For the
$s \rightarrow d \gamma$,
the summation is over the u,c,t quarks. The term proportional to
$m_q$ is small and usually neglected. The $F_1$ term does not
contribute to decays to real photons, however, it is relevant in weak
lepton-pair decays, like
$K^+ \rightarrow \pi^+ \ell^+ \ell^-$ $^{15,16}$,
$K^o_L \rightarrow \pi^o \ell^+ \ell^-$ $^{17,18}$;
it could also contribute, in principle, to leptonic decays of hyperons,
like
$\Sigma^+ \rightarrow p \ell^+ \ell^-$ $^{19}$
or
$\Omega^- \rightarrow \Xi^- \ell^+ \ell^-$ $^{20}$.

The QCD corrections are calculated by using an
operator product expansion combined with renormalization group
techniques. The effective Hamiltonian has the form
\begin{eqnarray}
H^{\Delta S=1}_{eff} = -
\frac{4G_F}{\sqrt{2}}
V_{CKM} \sum_i
c_i(\mu) O_i (\mu) \ .    
\end{eqnarray}
$c_i$
are Wilson coefficients and
$O_i$
are a set of renormalized local operators, generated by the electroweak
interactions and QCD. Physical amplitudes should be independent of the
renormalization scale $\mu$.
The relevant operators
in our case are$^{18}$
$O_{1,2}$,
the four QCD-penguin operators
$O_{3-6}$,
the electro- and chromomagnetic penguins
$O_{7 \gamma}$,
$O_{8G}$,
the electroweak penguins
$O_{7-10}$
and the semi-leptonic operators
$O_{9V}$,
$O_{10A}$.
We give here those employed here more frequently and we refer the
reader to Ref. 18 for the rest:
\setcounter{num}{3}
\setcounter{equation}{0}
\def\theequation{\thenum\alph{equation}}
\begin{eqnarray}
O_1 = (\bar{u}_{\alpha} \gamma_{\mu} P_L s_{\beta})
(\bar{d}_{\beta} \gamma^{\mu} P_L u_{\alpha});
O_2 = (\bar{u}_{\alpha} \gamma_{\mu} P_L s_{\alpha})
(\bar{d}_{\beta} \gamma^{\mu} P_L u_{\beta})    
\end{eqnarray}
\begin{eqnarray}
O_{7 \gamma} =
\frac{e}{16 \pi^2}
m_s
(\bar{d}_{\alpha} \sigma_{\mu\nu} P_R s_{\alpha}) F^{\mu\nu} ;
O_{8G} =
\frac{g_s}{16 \pi^2}
m_s
(\bar{d}_{\alpha} \sigma_{\mu\nu} T^a_{\alpha\beta}
P_R s_{\beta}) G^{\mu\nu}          
\end{eqnarray}
\begin{eqnarray}
O_{9V} =
(\bar{d}_{\alpha} \gamma_{\mu} P_L s_{\alpha})
(\bar{e} \gamma^{\mu} e) ;
O_{9A} =
(\bar{d}_{\alpha} \gamma_{\mu} P_L s_{\alpha})
(\bar{e} \gamma^{\mu} \gamma_5 e)               
\end{eqnarray}
\setcounter{equation}{3}
\def\theequation{\arabic{equation}}
where
$P_R,P_L$
are projection operators.

The Wilson coefficients are first computed perturbatively at
$\mu = M_W$
in zeroth order of QCD and then evolved down to the scale of the process
$\mu = m_0$
using the renormalization group equations (RGE). The solution of RGE
to leading logarithmic order is given in terms of effective coefficients
$c^{eff}_k (\mu = m_0)$,
so that one obtains
$c^{eff}_k (\mu = m_0) = u_{k\ell} c_{\ell} (M_W)$.

The function $F_1$ is related to
$c^{eff}_{9V}$
and
$F_2$
is related to
$c^{eff}_{7 \gamma}$.
In the next section, we shall discuss explicit calculations of the
contribution of the short-distance
$s \rightarrow d \gamma$
to several kaon decays.
\vspace{0.5cm}

\noindent {\bf 3. \ Short-Distance
$s \rightarrow d\gamma$
Contribution in Kaon Decays} \\

The decay which has been firstly considered in detail$^{15}$ as a
possible handle on the
$s \rightarrow d \gamma$
transition, with a virtual photon, is
$K^+ \rightarrow \pi^+ \ell^+ \ell^-$.
The decay amplitude can be written as
\begin{eqnarray}
M (K^+ \rightarrow \pi^+ \ell^+ \ell^-)
= \frac{ie^2G_F}{4\sqrt{2} \pi^2}
s_1c_1c_3 As \cdot
(p_K + p_{\pi})^{\mu}
\frac{\bar{u}(p_-)\gamma_{\mu} v(p_+)}{s}    
\end{eqnarray}
where
$s = (p_- +p_+)^2 = (p_K - p_{\pi})^2, \
s_1 = \sin \theta_1$, etc.
(CKM angles). A contains short-distance (SD)
and long-distance (LD) contributions.
The LD
includes \ (a) transitions
$K \rightarrow S \pi$
followed by
$S \rightarrow \gamma^{\ast}$,
where $S$ is a strange intermediate hadronic state, \ (b) contributions
involving a
``$K-\pi$''
transition with pole, nonpole and contact terms and \ (c) contributions
from strong penguin diagrams. These three classes can be calculated now
fairly reliably$^{16}$ and are all of the order 1. A similar result
obtains in a chiral perturbation approach$^{21}$. Moreover, the
experimental rate and spectrum$^{22}$ of
$K^+ \rightarrow \pi^+ e^+ e^-$
agrees well with the predictions of Refs. [16,21]. Turning to the
short-distance part,
$s \rightarrow d \ell^+ \ell^-$
contains contributions from the electromagnetic penguin
$s \rightarrow d \gamma^{\ast}$,
the
$Z^o$
penguin
$s \rightarrow d Z^{o \ast}$
and the
``W box''
diagram. The latter two may become significant in view of the heaviness
of the top quark. However, since
$\mid V^{\ast}_{ts} V_{td} \mid /
\mid V^{\ast}_{cs} V_{cd} \mid
\simeq 2 \times 10^{-3}$,
it turns out that the contributions from
$Z^o$-penguin and W-box are only a few percent of the main
short-distance part which is due to the c-quark in the electromagnetic
penguin loop.
One finds$^{16}$
$A_{SD} = 0.10$,
hence the SD
contribution to the rate is negligible and one
cannot hope to detect its effect in rate or spectrum. This is also
the reason for choosing (4) for the amplitude representation, despite
the fact that in principle there is an axial part from
$O_{9A}$.

There is, however, the possibility of detecting the short-distance
contribution in
$K^+ \rightarrow \pi^+ \mu^+ \mu^-$
from the interference of the one-photon LD part with the
axial vector part coming from the SD contribution of
$Z^o$-penguin and W-box diagrams. This induces$^{23}$ in this decay a
very small parity-violating longitudinal muon polarization asymmetry
$\Delta_{LR} = (\Gamma_R - \Gamma_L)/(\Gamma_R + \Gamma_L)$,
where
$\Gamma_R (\Gamma_L)$
is the rate of producing a right-(left)-handed
$\mu^-$.

The above discussion on
$K^+ \rightarrow \pi^+ \ell^+ \ell^-$
indicates that the
LD contributions prevent an easy access to the SD
ones. To reach these, one must turn to very rare decays, of which three
have received considerable  theoretical attention$^{18}$ in recent years
and are hopefully within reach of experimental detection$^{24}$ in the
not too distant future. The three decays are
$K^o_L \rightarrow \pi^o e^+ e^-$,
$K^+ \rightarrow \pi^+ \nu \bar{\nu}$,
$K^o_L \rightarrow \pi^o \nu \bar{\nu}$.
We shall not discuss these decays in detail here except for a few
general remarks, and we refer the reader to Refs. [18,24] for
reviews on these modes. The
$K^o_L \rightarrow \pi^o e^+ e^-$
decay contains$^{17}$
a direct CP-violating part due to short-distance diagrams, an indirect
CP-violating part induced via
$K_s \rightarrow \pi^o e^+ e^-$
and a CP-conserving contribution. The direct CP-violating contribution
is governed by the electromagnetic penguin operator
$O_{7V}$
as well as by
$O_{7A}$
and the strong penguin operators
$O_3 - O_6$.
It is expected to occur at a branching ratio of
$\sim 4 \times 10^{-12}$,
however, the other two contributions are estimated at comparable
figures$^{25}$. The decay
$K^+ \rightarrow \pi^+ \nu \bar{\nu}$
is the classic one for detecting SD
contributions from Z-penguin and W-box, in view of the
smallness of LD ones. Presently, after the next-to-leading
order QCD contributions have been considered$^{26}$, one expects$^{18}$
$0.6 \times 10^{-10} \leq Br
(K^+ \rightarrow \pi^+ \nu \bar{\nu})
\leq 1.5 \times 10^{-10}$.
The
$K^o_L \rightarrow \pi^o \nu \bar{\nu}$
is also dominated by short-distance diagrams$^{27}$ and proceeds via
direct CP-violation.
The theoretical analysis predicts$^{18}$
$10^{-11} \leq Br
(K^o_L \rightarrow \pi^o \nu \bar{\nu})
\leq 5 \times 10^{-11}$.

We now turn to the part of
$s \rightarrow d \gamma$
which occurs only in decays to real photons, namely the magnetic
transition
$F_2$
of Eq. (1). Firstly, with the normalization of Eqs. (1), (2),
$F_2 = 2c_7$.
The coefficient
$c_7^{eff}$
has been calculated recently by using the expressions of Buras et
al.$^{12}$, with their anomalous dimension matrix, which gives
$c^{eff}_7 (\mu = m_0)$
in terms of
$c_2(M_W), c_7(M_W), c_8(M_W)$.
The last two coefficients contribute very little, since they are
multiplied by
$V^{\ast}_{ts}V_{td}$.
Using
$\alpha_s (M_W) = 0.12$,
$\alpha_s (m_c) = 0.3$,
$\alpha_s (\mu = m_0) = 0.9$
one obtains$^{28}$
$F_2 = 0.12$
which we shall use here. Other calculations$^{3,29,30}$ give
respectively 0.16, 0.20 and 0.08 for
$F_2$,
using
slightly different values as input in the calculation. This gives
an indication on the sensitivity of
$c^{eff}_7$
to the input parameters, which in the leading log approximation is of
the order of at least 50\%.

The most frequent kaon  decay sensitive to
$F_2$
is
$K^+ \rightarrow \pi^+ \pi^o \gamma$ $^{30,31}$.
The general form of the amplitude for the direct decay is$^{31}$
\begin{eqnarray}
M (K^+ \rightarrow \pi^+ \pi^o \gamma) =
A \epsilon^{\mu \nu \sigma \tau} p^{(+)}_{\mu} p^{(0)}_{\nu} k_{\sigma}
\epsilon_{\tau} +
B[
(p^{(+)} \cdot \epsilon)
(p^{(0)} \cdot k)  -
(p^{(0)} \cdot \epsilon)
(p^{(+)} \cdot k)] \ . 
\end{eqnarray}
It has a parity-conserving magnetic transition $A$ and a
parity-violating electric transition $B$. The former appears to be the
dominant contribution and the SD
$s \rightarrow d \gamma$
contributes to it via
$F_2$.
The calculation of Ref. 30 shows that this contribution, again, is only
a few percent of the LD contributions$^{31}$, which reproduce
the experimental findings.
Hence, here again, the branching ratio or the $\gamma$-spectrum cannot
be used to reach the
$s \rightarrow d \gamma$
transition.
\vspace{0.5cm}

\noindent {\bf 4. \ Dynamics of Hyperon Radiative Decays} \\

Since the more frequent radiative kaon decays are not useful for
investigating the SD
$s \rightarrow d \gamma$
transition, we turn to the baryonic (s,d,u) sector, in particular,
the hyperon radiative decays. These decays, despite intensive attention
during nearly thirty years, are still plagued with notorious
discrepancies between theoretical models and experiment (see Ref. 32
for a recent review). The general form of the amplitude for such decays
is
\begin{eqnarray}
(H \rightarrow H^{\prime} \gamma) = i e G_F \bar{u}(p^{\prime})
\sigma_{\mu \nu} (A + B \gamma_5) \epsilon^{\mu} k^{\nu} u(p)   
\end{eqnarray}
where
$H, H^{\prime}$
are spin
$\frac{1}{2}$
baryons. As such, it is an ideal place for investigating the $F_2$
transition of Eq. (1).

For our purpose here, we note that the seven weak radiative decays
which can occur within the octet and decuplet, can be divided$^{33}$
from a dynamical point of view into two groups, the ``pole decays''
($\Sigma^+ \rightarrow p \gamma, \
\Lambda \rightarrow n \gamma, \
\Xi^o \rightarrow \Sigma^o \gamma, \
\Xi^o \rightarrow \Lambda^o \gamma$)
and the ``non-pole decays''
($\Xi^- \rightarrow \Sigma^- \gamma, \
\Omega^- \rightarrow \Xi^- \gamma, \
\Omega^- \rightarrow \Xi^{\ast -} \gamma$).
The contribution of the SD
$s \rightarrow d \gamma$
to the ``pole decays'', occurring with branching ratios of the order
of
$10^{-3}$, is negligible$^{34,35}$. The second group of decays involves
particles
$\Omega^- (sss), \ \Xi^- (ssd), \ \Sigma^- (sdd)$
with no u-valent quark, hence there are no W-exchange diagrams to
generate poles. These decays may proceed via two-hadron intermediate
states$^{34,36}$; since they are expected to
have fairly low branching ratios
($10^{-4} - 10^{-5}$),
it has been suggested$^{35}$
that the contribution of the SD
$s \rightarrow d \gamma$
in the rate or in the asymmetry of decay might be detectable in some
of them, especially
$\Omega^- \rightarrow \Xi^- \gamma$.

So far, there are two measurements of the
$\Xi^-$-decay, giving$^{32}$
$Br(\Xi^- \rightarrow \Sigma^- \gamma) = 1.27 \pm 0.23 \times 10^{-4}$.
A calculation$^{36}$ using the intermediate state of
($\Lambda \pi$) only, gives a branching ratio
$Br^{(th)}
(\Xi^- \rightarrow \Sigma^- \gamma) = (1.8 \pm 0.4) \times 10^{-4}$
and an asymmetry coefficient
$^{(th)} \alpha (\Xi^- \rightarrow \Sigma^- \gamma) = - 0.13 \pm
0.07$.
The uncertainties are due mainly to the possible additional contribution
of
$s \rightarrow d \gamma$.
On the other hand, for
$\Omega^-$
decay there is only an upper limit$^{37}$
$Br (\Omega^- \rightarrow \Xi^- \gamma) < 4.6 \times 10^{-4}$.
\vspace{0.5cm}

\noindent {\bf 5. \ Short-Distance and Long-Distance
$s \rightarrow d \gamma$
Contributions in
$\Omega^- \rightarrow \Xi^- \gamma$} \\

In this section we concentrate on the
$s \rightarrow d \gamma$
contribution to the ``non-pole'' group of hyperon decays, in particular
to
$\Omega^- \rightarrow \Xi^- \gamma$.
In addition to the SD contribution$^{28,29}$,
there is also$^{28}$
a long-distance contribution of
$s \rightarrow d \gamma$,
which will be analyzed here.

Firstly, we note that the
$s \rightarrow d \gamma$
transition (whether short- or long-distance) cannot be the dominant
transition in {\it all} hyperon radiative decays. Indeed, if one
assumes$^5$ that its strength can be derived from the observed
$\Sigma^+ \rightarrow p\gamma$
rate, the other radiative decays are predicted to be much larger than
observed.

The SD amplitude for
$s \rightarrow d \gamma$
is given by the
$F_2$
term in Eq. (1). For the LD one, we consider a vector meson dominance
(VMD) approximation to compute
$s \rightarrow dV \rightarrow  d\gamma$
transitions, along the lines discussed recently$^{38}$ for
$b \rightarrow s \gamma$.
We remark that a long-distance
$s \rightarrow d \gamma$
transition is also one of the various contributions in
$K^+ \rightarrow \pi^+e^+e^-$,
as considered long ago by Vainshtein et al.$^{15}$ with a somewhat
different technique. Using the operators
$O_1, O_2$
with u and c quarks and factorization, one obtains$^{38}$ for the
amplitude
$s \rightarrow d + V$
\begin{eqnarray}
A(s \rightarrow d + V_i) =
ig_{V_i}(q^2)
\frac{G_F}{\sqrt{2}} a_2 V_{is} V^{\ast}_{id} \bar{d} \gamma_{\mu}
(1-\gamma_5) s \epsilon^{\mu^+}    
\end{eqnarray}
where
$a_2$
is a QCD correction factor and
$g_{V_i}(q^2)$
is defined
$<V_i (a) \mid \bar{q}_i \gamma_{\mu} q_i \mid 0 > =
ig_{V_i}(q^2)
\epsilon^+_{\mu}$.
After extracting the transverse part of the amplitude in (7) to couple
it to photons, the
$s \rightarrow d \gamma$
amplitude including both SD and LD  contributions is given by$^{28}$
\begin{eqnarray}
&& A (s \rightarrow d \gamma)
= A_{SD} + A_{LD} = -
\frac{eG_F}{\sqrt{2}} \bar{d} \sigma^{\mu\nu}
[(\frac{m_sF_2}{8\pi^2} +
\frac{va_2C_{VMD}M_s}{M^2_s-M^2_d})P_R \nonumber \\
&& +(\frac{m_dF_2}{8\pi^2} -
\frac{va_2C_{VMD}M_d}{m^2_s-M^2_d}) P_L] s F_{\mu v}  
\end{eqnarray}
where
$v = \mid V^{\ast}_{cs} V^{\ast}_{cd} \mid 0.22$,
$a_2 \simeq 0.5, \ m_s, \ m_d$
are current masses and
$M_s, \ M_d$
are constituent quark masses.
\begin{eqnarray}
C_{VMD} = \frac{2}{3} \sum_i
\frac{g^2_{\psi_i}(0)}{m^2_{\psi_i}} -
\frac{1}{2}
\frac{g^2_{\rho}(0)}{m^2_{\rho}} -
\frac{1}{6}
\frac{g^2_{\omega}(0)}{m^2_{\omega}}  
\end{eqnarray}
with the summation over the six narrow $c\bar{c}$ resonances.
To calculate the physical decay
$\Omega^- \rightarrow \Xi^- \gamma$
one uses$^{28}$ the amplitude (8) with SU(6) quark-model wave functions
for the baryons and unit overlap. If we use only
$A_{SD}$,
with
$F_2 = 0.12$,
one finds
$\Gamma_{SD} (\Omega^- \rightarrow \Xi^- \gamma) \simeq 6.4 \times
10^{-12}$ eV,
far below the present experimental bound$^{37}$ of
$3.7 \times 10^{-9}$ eV.
Using a value of
$F_2 = 0.2$
and QCD sum rules to estimate the contribution of
$A_{SD}$
to the
$\Omega^- \rightarrow \Xi^- \gamma$
decay, Nielsen et al.$^{29}$ find a value of
$\Gamma_{SD}$
which is about 8 times larger,  but still two orders of magnitude below
the experimental limit.

Using the full expression (8), the experimental upper limit determines
$\mid C_{VMD} \mid < 0.01$ GeV$^2$.
Assuming
$g_V(0) \simeq g_V(m^2_V)$
for
$V = \rho, \omega$,
one obtains from (9)
$\sum_i
\frac{g^2_{\psi_i}(0)}{m^2_{\psi_i}}
= 0.045 \pm 0.016$ GeV$^2$,
about one sixth of the value at
$m^2_{\psi_i}$.
This confirms previous estimates summarized in Ref. 38 and shows that
the right hand side of Eq. (9) is a remarkable sum rule holding to about
30\%. This stems from the underlying input, which essentially include
the GIM mechanism and
SU(4)$_{\mbox{F}}$
symmetry.

It is clear from the above discussion that the LD contribution to
$s \rightarrow d \gamma$
may be significantly larger than the SD one and might even saturate
the present experimental limit. It is important at this stage to mention
additional possible contributions to the
$\Omega^- \rightarrow \Xi^- \gamma$.
Firstly, there is the unitarity limit, giving$^{34}$
Br$^{\mbox{unit}}
(\Omega^- \rightarrow \Xi^- \gamma)
> 0.8 \times 10^{-5}$.
The dominant two-body intermediate state of
$\Xi^o \pi^-$
contributes$^{34}$
$(1-1.5) \times 10^{-5}$
to the branching ratio and the strong penguin contribution is$^{39}$
about
$4 \times 10^{-6}$
or less. Hence, if the decay will turn out to be between the present
upper limit of
$4.6 \times 10^{-4}$
and down to about
$3 \times 10^{-5}$,
the LD t-channel contribution of
$s \rightarrow dV \rightarrow d\gamma$
is the dominant one. In this case the asymmetry of the decay should
be$^{40}$
$\alpha (\Omega^- \rightarrow \Xi^- \gamma) = 0.4 \pm 0.1$,
vs., say, the SD contribution which leads to an asymmetry 1. One should
also add that the LD contribution of
$s \rightarrow d \gamma$
with
$\mid C_{VMD} \mid \leq 0.01$,
is consistent with the observed hyperon decays$^{28}$.

In concluding, we stress that the more frequent kaon radiative decays
are not profitable terrain for investigating the
$s \rightarrow d \gamma$
transition and one must pursue the detection of the very rare decays,
like
$K^+ \rightarrow \pi^+ \nu \bar{\nu}$,
$K^o_L \rightarrow \pi e^+ e^-$,
$K_L \rightarrow \pi^o \nu \bar{\nu}$
in order to check the SD diagrams. On the other hand, the
$\Omega^- \rightarrow \Xi^- \gamma$
decay might provide very useful information on the SD/LD ratio in
$s \rightarrow d \gamma$.
In fact, from (8) and the experimental upper limit on
$\Omega^- \rightarrow \Xi^- \gamma$
we already know that
(LD/SD) $\leq 25$
in this decay. Thus
$s \rightarrow d \gamma$
represents an intermediate situation between the SD dominated
$b \rightarrow s \gamma$,
and
$c \rightarrow u \gamma$
decays where the SD contribution is negligible.
\vspace{0.5cm}

\noindent {\bf References}
\baselineskip 14pt plus 2pt
\begin{enumerate}
\item M.K. Gaillard and B.W. Lee,
Phys. Rev. {\bf D10}, 897 (1974).
\vspace{-0.25cm}
\item T. Inami and C.S. Lim,
Prog. Theor. Phys. {\bf 65}, 297 (1981).
\vspace{-0.25cm}
\item M.A. Shifman, A.I. Vainshtein and V.I. Zacharov,
Phys. Rev. {\bf D18}, 2583 (1978).
\vspace{-0.25cm}
\item M.A. Ahmed and G.G. Ross,
Phys. Lett. {\bf 59B}, 293 (1975); \\
N. Vasanti,
Phys. Rev. {\bf D13}, 1889 (1976).
\vspace{-0.25cm}
\item F.J. Gilman and M.B. Wise,
Phys. Rev. {\bf D19}, 976 (1979).
\vspace{-0.25cm}
\item B.A. Campbell and P.J. O'Donnell,
Phys. Rev. {\bf D25}, 1989 (1982).
\vspace{-0.25cm}
\item S. Bertolini, F. Borzumati and A. Masiero,
Phys. Rev. Lett. {\bf 59}, 180 (1987);
N.G. Deshpande, P. Lo, J. Trampetic, G. Eilam and P. Singer,
Phys. Rev. Lett. {\bf 59}, 183 (1987).
\vspace{-0.25cm}
\item B. Grinstein, R. Springer and M.B. Wise,
Nucl. Phys. {\bf B339}, 269 (1990);
M. Ciuchini et al.,
Phys. Lett. {\bf B316}, 127 (1993);
M. Misiak,
Nucl. Phys. {\bf B393}, 23  (1993);
{\bf B439}, 461 (1995) (E);
G. Cella et al.,
Nucl. Phys. {\bf B431}, 417 (1994).
\vspace{-0.25cm}
\item M. Misiak and M\"{u}nz,
Phys. Lett. {\bf B344}, 308 (1995).
\vspace{-0.25cm}
\item M.S. Alam et al.,
Phys. Rev. Lett. {\bf 74}, 2885 (1995).
\vspace{-0.25cm}
\item R. Ammar et al.,
Phys. Rev. Lett. {\bf 71}, 674 (1993).
\vspace{-0.25cm}
\item A.J. Buras, M. Misiak, M. M\"{u}nz and S. Pokorski,
Nucl. Phys. {\bf B424}, 374 (1994).
\vspace{-0.25cm}
\item A. Ali and C. Greub,
Z. Phys. {\bf C60}, 433 (1993).
\vspace{-0.25cm}
\item G. Burdman, E. Golowich, J.L. Hewett and S. Pakvasa,
Phys. Rev. {\bf D52}, 6383 (1995).
\vspace{-0.25cm}
\item A.I. Vainshtein et al.,
Yad. Fiz. {\bf 24} (1976);
F.J. Gilman and M.B. Wise,
Phys. Rev. {\bf D21}, 3150 (1980).
\vspace{-0.25cm}
\item L. Bergstr\"{o}m and P. Singer,
Phys. Rev. Lett. {\bf 55}, 2633 (1985).
Phys. Rev. {\bf D43}, 1568 (1991).
\vspace{-0.25cm}
\item C.O. Dib, I. Dunietz and F.J. Gilman,
Phys. Rev. {\bf D39}, 2639 (1989).
\vspace{-0.25cm}
\item G. Buchalla, A.J. Buras and M.E. Lautenbacher,
Rev. Mod. Phys. (1996).
\vspace{-0.25cm}
\item S.-P. Chia and G. Rajagopal,
Phys. Lett. {\bf B156}, 405 (1985);
L. Bergstr\"{o}m, R. Safadi and P. Singer,
Z. Phys. {\bf C37}, 281 (1988).
\vspace{-0.25cm}
\item R. Safadi and P. Singer,
Phys. Rev. {\bf D37}, 697 (1988);
{\bf 42}, 1856 (E) (1990).
\vspace{-0.25cm}
\item G. Ecker, A. Pich and E. De Rafael,
Nucl. Phys. {\bf B291}, 692 (1987);
{\bf B303}, 665 (1988).
\vspace{-0.25cm}
\item C. Alliegro et al.,
Phys. Rev. Lett. {\bf 68}, 278 (1992).
\vspace{-0.25cm}
\item M. Savage and M. Wise,
Phys. Lett. {\bf B250}, 151 (1990);
\vspace{-0.25cm}
\item L. Littenberg and G. Valencia,
Ann. Rev. Nucl. Part. Sci. {\bf 43}, 729 (1993).
\vspace{-0.25cm}
\item P. Heiliger and L. Sehgal,
Phys. Rev. {\bf D47}, 4920 (1993).
\vspace{-0.25cm}
\item G. Buchalla and A.J. Buras,
Nucl. Phys. {\bf B412}, 106 (1994).
\vspace{-0.25cm}
\item L. Littenberg,
Phys. Rev. {\bf D39}, 3322 (1989).
\vspace{-0.25cm}
\item G. Eilam, A. Ioannissian, R.R. Mendel and P. Singer,
Phys. Rev. {\bf D53}, 3629 (1996).
\vspace{-0.25cm}
\item M. Nielsen et al.,
Phys. Rev. {\bf D53}, 3620 (1996).
\vspace{-0.25cm}
\item M. McGuigan and A.I. Sanda,
Phys. Rev. {\bf D36}, 1413 (1987).
\vspace{-0.25cm}
\item M. Moshe and P. Singer,
Phys. Lett. {\bf 51B}, 367 (1974); H.-Y. Cheng,
Phys. Rev. {\bf D49}, 3771 (1994).
\vspace{-0.25cm}
\item Y. Lach and P. Zenczykowski,
Inst. J. Mod. Phys. {\bf A10}, 3817 (1995).
\vspace{-0.25cm}
\item P. Singer,
Nucl. Phys. B (Proc. Suppl.)  (1996).
\vspace{-0.25cm}
\item Ya. I. Kogan and M.A. Shifman,
Yad. Fiz. {\bf 38}, 1045 (1983).
\vspace{-0.25cm}
\item L. Bergstr\"{o}m and P. Singer,
Phys. Lett. {\bf B169}, 297 (1986).
\vspace{-0.25cm}
\item P. Singer,
Phys. Rev. {\bf D42}, 3255 (1990).
\vspace{-0.25cm}
\item I.F. Albuquerque et al.,
Phys. Rev. {\bf D50}, R18 (1994).
\vspace{-0.25cm}
\item N.G. Deshpande, X.-G. He and J. Trampetic,
Phys. Lett. {\bf B367}, 362 (1996).
\vspace{-0.25cm}
\item S.G. Kamath,
Nucl. Phys. {\bf B198}, 61 (1982); J.O. Eeg, z. Phys. C {\bf 21}, 253
(1984).
\vspace{-0.25cm}
\item G. Eilam, A. Ioannissyan and P. Singer,
TECHNION-PH-96-7.
\end{enumerate}

\end{document}